\documentclass[prl,reprint,
superscriptaddress,
amsmath,amssymb,aps]{revtex4-2}

\usepackage[T1]{fontenc}
\usepackage[english]{babel}
\usepackage{hyperref}
\usepackage{graphicx}
\usepackage{subcaption}
\usepackage{siunitx}
\usepackage{titlesec}
\usepackage{upgreek}
\usepackage{xcolor}
\usepackage{bm}

\newcommand{\dox}[0]{$\mathrm{D^0X}$}
\newcommand\LCN{\,London Centre for Nanotechnology, UCL, 17-19 Gordon St, London WC1H 0AH, UK}
\newcommand\EEE{\,Department of Electrical and Electronic Engineering, UCL, Malet Place, London WC1E 7JE, UK}
\newcommand\IITM{\,Department of Physics, Indian Institute of Technology Madras, Chennai 600036, India}

\begin{document}

\title{Donor-bound-exciton strain microscopy in silicon devices}

\author{Pierandrea Conti}
\altaffiliation{pierandrea.conti.16@alumni.ucl.ac.uk}
\affiliation{\LCN}
\author{Siddharth Dhomkar}
\affiliation{\LCN}
\affiliation{\IITM}
\author{Philipp Ross}
\affiliation{\LCN}
\author{John Mansir}
\affiliation{\LCN}
\author{John J. L. Morton}
\altaffiliation{jjl.morton@ucl.ac.uk} \affiliation{\LCN}
\affiliation{\EEE}

 \begin{abstract}
We explore the effects of stress on silicon donor bound exciton (\dox{}) transitions in bulk silicon and in microfabricated silicon devices.
We first study \dox{} transitions in an isotopically purified silicon-28 bulk doped  sample under controlled uniaxial stress, confirming the validity of existing models in the low strain ($\lesssim 10^{-5}$) regime.
We then demonstrate the localised photoconductive detection of a few thousand donors illuminated by a 1078~nm resonant laser with \qty{4}{\micro\meter} spot focused on a microfabricated device consisting of an implanted phosphorus layer between a pair of metallic contacts. We observe local variations in the strained exciton peak splitting from \qtyrange{10}{200}{\micro\electronvolt}, and obtain scanning microscopy stress maps in good agreement with finite-element-model thermal stress simulations. Our results suggest a potential use of donor bound excitons for in-situ stress sensing, and demonstrate pathways for the miniaturisation of \dox{} photoconductive detection.
\end{abstract}

\maketitle

\section{Introduction}

Strain is an integral part of semiconductor quantum devices, whether it has been intentionally engineered through the growth of a heterostructure or arises from stresses at cryogenic operating temperatures due to differing thermal expansion coefficients of the device materials~\cite{hu91,thorbeck15,yu10}. 
Electron spins within silicon, which have long been studied as qubit candidates, can be influenced by strain, for example through its effect on the hyperfine interaction for donor spins~\cite{koiller02,mansir18,ranjan21}. However, recent results have stimulated rapidly growing interest in \emph{hole} spin qubits in silicon and germanium, demonstrating rapid high-fidelity electrical qubit control~\cite{lawrie23,liles23}, multi-qubit operations~\cite{hendrickx21} and spin-photon coupling~\cite{yu23}. For such hole spins, whether bound to acceptors or in quantum dots, the impact of strain is profound due to the large spin-orbit coupling~\cite{kobayashi21,piot22,hendrickx24}. Understanding how strain is distributed throughout complex semiconductor devices, and the influence of device design and materials, is therefore critical in order to develop technologies based on such spin systems. 

Donor bound excitons (\dox) in silicon have been studied as a promising pathway to the opto-electronic readout of donors spins, thanks to their narrow optical transitions with homogeneous linewidths as low as \qty{20}{\nano\electronvolt}\cite{yang09,karaiskaj01}. Spin-selective optical generation of the \dox, followed by Auger recombination\cite{schmid77} yielding $\mathrm{D^0 \rightarrow D^+}$ provides an effective method for donor spin-to-charge conversion, which has been demonstrated in isotopically purified silicon\cite{lo15,franke16,ross19}. 
However, donor bound exciton transitions are sensitive to strain due to the conduction band valley orbit\cite{wilson61} and valence band Pikus-Bir interactions\cite{bir63}, yielding line shifts, splitting and broadening.  The effects of strain in \dox\ transitions have been explored by Thewalt \emph{et al.} under compressive uniaxial stress as high as \qty{200}{\mega\pascal}~\cite{thewalt78}, as well as by Lo~\emph{et al.}~\cite{lo15} and Loippo~\emph{et al.}\cite{loippo23}, who observed strain split peaks due to the interface strain from metal contacts and epitaxial lattice mismatch. Splitting and broadening of the \dox{} peaks have been studied to assess lattice strain, analysing the residual ion implantation lattice damage in bulk~\cite{peach18,peach19} and SOI~\cite{sumikura11} silicon samples.

The strain sensitivity of the \dox\ transition provides a potential route towards local opto-electronic measurements of strain in silicon nanodevices. So far, however, \dox{} photoconductive readout has been limited to bulk crystals or  relatively large bulk doped devices ($700\times100~\upmu{\rm m}^2$ in area) and illumination areas far above the diffraction limit\cite{loippo23,lo15}.  
In this work, we demonstrate how \dox\ transitions in silicon nanodevices can be studied using scanning optical microscopy, using the position-dependent spectra to obtain equivalent stress images of the device, with a laser spot size of \qty{4}{\micro\meter} illuminating fewer than \num{10000} donors. We begin with a study of \dox\ transitions in the low strain regime using the narrow peaks of an isotopically purified $\mathrm{^{28}Si}$ sample, in order to establish the model and its parameters which enable equivalent stress to be determined from the optical spectra. We then demonstrate the detection of donor bound excitons in strained, implanted phosphorus layers in microfabricated devices, and explore their spectral characteristics with a high-numerical aperture scanning microscopy setup, confirming the obtained images with finite element model (FEM) simulations. At this scale, the impacts of strain inherent to microfabricated devices are evident, especially those arising from thermal interface stress. In addition to providing a powerful probe of local strain within silicon nanodevices, these results may help to overcome challenges which have so far hampered efforts to extend \dox{}-based spin readout to nanoscale devices~\cite{mansir18} and SOI substrates~\cite{sumikura11}.

\section{Donor Bound Excitons in the Low Strain Regime}
\begin{figure}[t]
	\centering
\includegraphics[width=\linewidth]{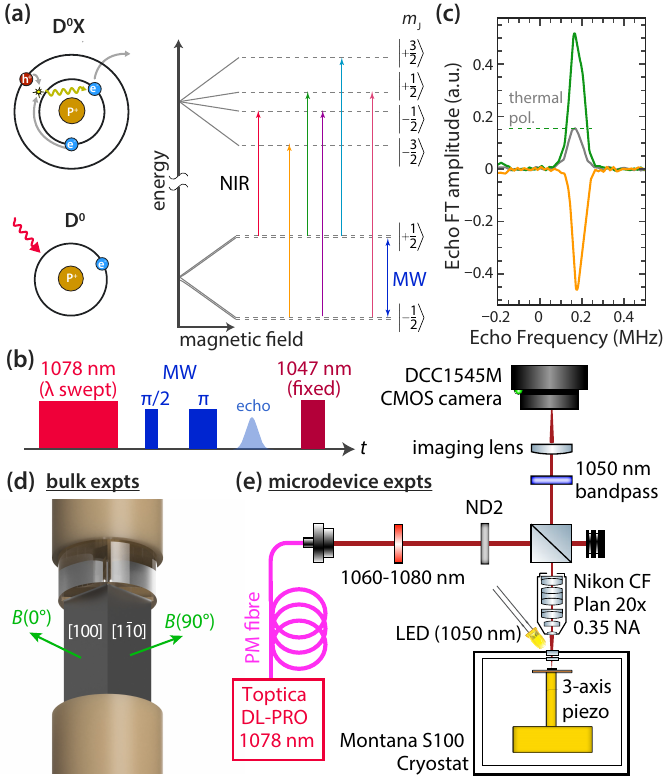}
	\caption{{\bf The \dox{} energy transitions of P donors in silicon and experimental set-ups.} (a) Six allowed optical transitions in the near infrared (NIR, 1078 nm) can be readily resolved, detected via their influence on electron spin resonance (ESR) experiments performed using the microwave (MW) transition on the neutral donor. A Hahn echo sequence (b) is used to study the effect of optical illumination on donor spin polarisation.
 (c) Measurements from the bulk sample in the dark provide a signal level (grey) corresponding to thermal spin polarisation. By optically exciting a \dox{} transition, the donor spin may be positively (green) or negatively (orange) hyperpolarised.
 (d) CAD rendering of the silicon crystal sample and holder for stress application used in bulk measurements. (e) Optical set-up used in the scanning measurements of microfabricated silicon devices, in which \dox\ spectra are obtained by monitoring impedance changes in the device. }\label{fig:rod}
\end{figure}

The donor bound exciton transition excites a neutral donor $\mathrm{D^0}$ with spin number $|S=\frac{1}{2}, m_S=\pm \frac{1}{2}\rangle$ into a three-particle $\mathrm{D^0X}$ state with spin number determined by the hole $|S=\frac{3}{2}, m_S=\pm \frac{1}{2}, \pm \frac{3}{2}\rangle$ state, leading to 8 spin dependent transitions of which two are forbidden by dipole selection rules, as shown in Fig.~\ref{fig:rod}(a). Strain shifts the transition energies via the deformation potential, the conduction band valley-orbit interaction~\cite{wilson61} and the valence band Pikus-Bir interaction~\cite{bir63},
\begin{equation} \label{eq:pikus_bir}
\hat{H}_\mathrm{pb}\left(\epsilon_{ij}\right) = a^\prime \mathrm{Tr}\{\epsilon\}+b^\prime \sum_i(\hat{J_i}^2-\frac{5}{4}\hat{I})\epsilon_{ii} + \frac{2d^\prime}{\sqrt{3}}\sum_{j>i}\left\lbrace  \hat{J_i}\hat{J_j} \right\rbrace \epsilon_{ij}.
\end{equation}
The latter introduces a spin-lattice interaction and thus a strain induced spin splitting, much like for the hole spin states of acceptors in silicon.

We apply strain via a pair of machined polyetheretherketone rods (see Figs.~\ref{fig:rod}(d)) enclosing a $\mathrm{^{28}Si}$ sample with dimensions $2\times2\times7~\mathrm{mm^3}$, doped with phosphorus at a concentration of \qty{1.5e14}{\per\cubic\centi\meter}.
Calibrated masses are added to a platform mounted on the top rod, ouside the cryostat, to apply uniaxial compression stress to the sample, as described in Ref.~\cite{mansir18} and in Methods.
We use X-band pulsed electron spin resonance (ESR) to observe the \dox\ transitions through their effect on the P-donor electron spin polarisation, as shown in Fig.~\ref{fig:rod}(c) (see Methods). Measurements are performed at \qty{4.2}{\kelvin}, with laser illumination provided through the cryostat window. The pulse sequence, illustrated in Fig.~\ref{fig:rod}(b), consists of a \qty{200}{\milli\second} pulse from a \qty{1078}{\nano\meter} Koheras Adjustik tunable laser to polarise the donor spins, followed by an ESR Hahn echo sequence for polarisation detection and a \qty{20}{\milli\second} above-bandgap (\qty{1047}{\nano\meter}) laser pulse to reset the spin polarisation and reneutralise donors for the next experiment. 
The donor spin polarisation is derived from the integral of the Hahn echo in the Fourier space in order to minimise the effect of ESR frequency shifts induced by the stress. 
The (unstrained) \dox\ spectra indicate an optical hyperpolarisation of up to \qty{\sim 15}{\percent}, assuming a thermal equilibrium polarisation of \qty{5.1}{\percent} at 4.2~K and $B_0\sim0.35$~T. This is significantly less than the $>$\qty{90}{\percent} hyperpolarisation previously demonstrated~\cite{yang09a,lo15, morse18, ross19}, likely due to a portion of the sample being concealed by the rods, while still contributing to the magnetic ESR signal.

\begin{figure}[t]
    \centering
\includegraphics[width=0.95\linewidth]{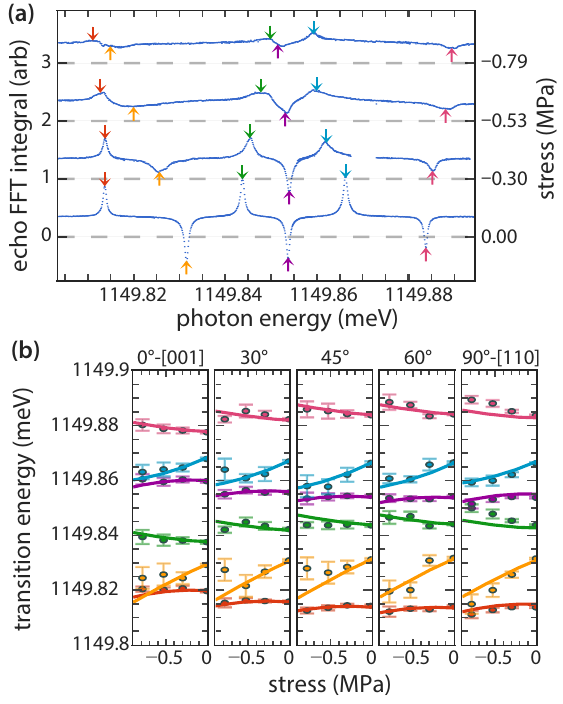}
	\caption{{\bf \dox{} transition energy shifts under compressive stress in a bulk doped $\mathrm{^{28}Si}$ sample.} (a) \dox{} spectra measured under variable compressive stress, obtained by an ESR Hahn Echo measurement, with magnetic field $B_0=345~\mathrm{mT}$ and ESR frequency $f_0=9.7~\mathrm{GHz}$. (b) Angle dependent energy shifts for each of the six transitions are extracted from the spectra and fit to a theoretical model (see text).} \label{fig:bulk}
\end{figure}

\dox{} hyperpolarisation spectra, such as the ones shown in Fig.~\ref{fig:bulk}(a) are obtained by repeating the Hahn echo pulse sequence while sweeping the wavelength of the first laser pulse, enabling a study of the effects of sample stress and magnetic field orientation. The \dox\ spectra display strain-induced energy shifts, evident particularly from the heavy hole transitions, together with broadening of the peaks, possibly due to some inhomogeneity in the applied strain. 
The six resolved \dox{} transition energies are extracted by fitting each hyperpolarisation spectrum to six Lorentzian lineshapes, in order to determine their dependence on field orientation and stress, as shown in Fig.~\ref{fig:bulk}(b). Theoretical transition energies shifts are derived from the difference between the donor bound exciton hole and donor electron energy levels. The \dox{} hole spin energy levels are given by the eigenvalues of the Zeeman and Pikus-Bir Hamiltonians  $\hat{H}_\mathrm{h}=\hat{H}_\mathrm{hZ} + \hat{H}_\mathrm{pb}$, where $H_\mathrm{hZ} =  \sum_{i=x,y,z}\left( g_1\hat{J_i}B_i+g_2\hat{J_i}^3B_i\right)$ \cite{bhattacharjee72,lo15}. The donor electron spin energy levels are given by the Zeeman Hamiltonian $\hat{H}_\mathrm{e}=g_e\mu_B B_0\hat{S_z}$, specifically the two levels driven by the ESR transition. There are further strain terms in the conduction band Hamiltonian due to the valley-orbit interaction. However, for a single strain orientation and low strains these cannot be distinguished from the Pikus-Bir diagonal term, and are thus introduced via a modified $a^{\prime\prime}$ parameter.
These theoretical transition energies are then used to fit the experimental transitions, yielding the parameters presented in Table~\ref{tab:parameters}.

The derived g-factor and Pikus-Bir parameters are in agreement with previous experimental results~\cite{thewalt78,lo15} and theoretical values from band structure calculations~\cite{yu10,blacha84}, and our results confirms the validity of the Pikus-Bir model in the low strain regime. 
The slightly lower values obtained in this work could be explained by the presence of some friction between the resonator mount and the rod lowering the effective strain applied to the sample. 

\begin{table}[t]
	\footnotesize
	\centering
	\begin{tabular}{|>{\centering}p{0.24\linewidth}|>{\centering}p{0.16\linewidth}|>{\centering}p{0.19\linewidth} >{\centering}p{0.14\linewidth} >{\centering\arraybackslash}p{0.12\linewidth}|}
		\hline
		& This Work & Lo\cite{lo15}, Thewalt\cite{thewalt78} & Loippo\cite{loippo23} & theory \cite{yu10,blacha84} \\
		\hline\hline
		$B \text{ (mT)}$ & $345(1)$ & $0$ & $0\to 300$ & \\
		$\sigma \text{ (MPa)}$ & $0 \to -1$ & $0 \to -150$ & $\sim 1$ & \\
		\hline\hline
		$g_\mathrm{e}$ & $2.01(2)$ & $1.9985$ & & \\
		$g_1$ & $-0.83(2)$ & $-0.8$ & $-0.83$ & \\
		$g_2$ & $-0.215(11)$ & $-0.24$ & $-0.22$ & \\
		$b' \text{ (eV)}$ & $-1.5(2)$ & $-1.7$  & $-7$ & $-2.2 $ \\
		$d' \text{ (eV)}$ & $-4.2(3)$ & $-5.1$ & $-4$ &  $-5.2$ \\
		\hline
	\end{tabular}
	\caption{Parameters of the strained \dox{} Hamiltonian extracted from bulk crystal measurements including a comparison with theory and literature values.} \label{tab:parameters}
\end{table}

\section{\dox{} Device Scanning Microscopy}

By establishing the validity of the Pikus-Bir model in the low strain regime, we have obtained an approach to translate predicted stress and strain maps within microfabricated devices into predictions for sample position-dependent \dox{} spectra. Conversely, it is possible to convert experimental peak-splittings into scalar measures of distortion stress, employing the bulk derived $b'$, $d'$ parameters as well as the compliance matrix terms $s_{11}$, $s_{12}$ and $s_{44}$. 

\begin{align} \label{eq:splitting}
\begin{split} 
    \Delta E(\bm{\sigma})  = & \, 2\bigg\{ \frac{1}{2}  b'^2(s_{11}-s_{12})^2 \Big[(\sigma_{xx}-\sigma_{yy})^2+(\sigma_{yy}-\sigma_{zz})^2 \\ & + (\sigma_{zz}-\sigma_{xx})^2 \Big] + \frac{d'^2s_{44}^2}{4}\big(\sigma_{xy}^2+\sigma_{yz}^2+\sigma_{zx}^2 \big) \bigg\}^\frac{1}{2}
\end{split}
\end{align}

The peak splitting described in Eq.~\ref{eq:splitting} closely resembles the expression for Von Mises equivalent stress, $\sigma_v$, a scalar measure combining principal and shear stresses with equivalent distortion energy. Specifically, the ratio of the shear and principal stress parameters closely resembles the 6 of $\sigma_v$(see  Eq. \ref{eq:von_mises}), with $\frac{d^2s_{44}^2}{2b^2(s_{11}-s_{12})^2}\approx 6.4$, consistent with a nearly isotropic \dox{} peak splitting under principal and shear stresses. We use this to obtain an approximate value of equivalent stress from the \dox{} peak splitting ($\sigma_v \approx C \Delta E$), with $C=34(3)\,\mathrm{GPa\,eV^{-1}}$ (see Methods).

\begin{figure}[t]
	\centering
\includegraphics[width=\linewidth]{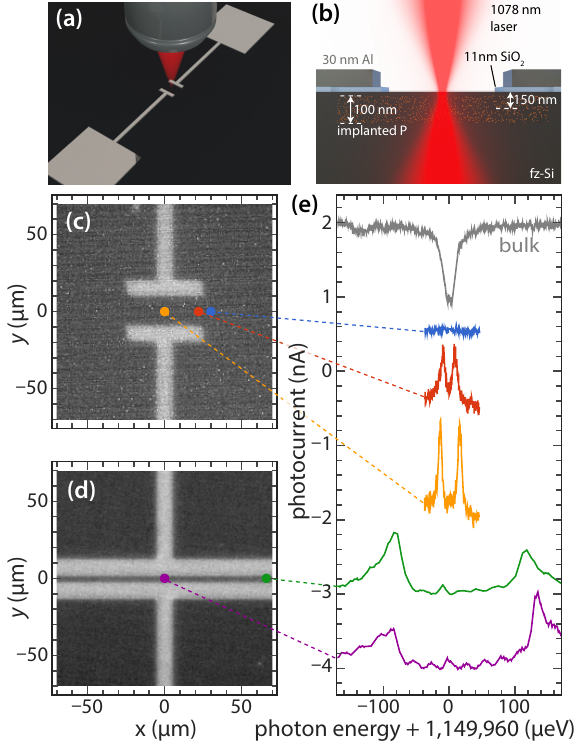}
	\caption{{\bf Position-dependent \dox{} spectra measured using the photocurrent in silicon devices}. (a,b) Illustrations of the device arrangement and materials composition. 
 (c,d)~Optical images of two devices studied, where the region between the MOS contacts has dimensions ($20 \times 50~\upmu{\rm m}^2$) and ($5 \times 400~\upmu{\rm m}^2$), respectively. (e)~Photocurrent spectra (vertically offset for clarity) measured on the devices at the positions indicated. The (\textcolor{gray}{\rule[0.5ex]{15pt}{1.5pt}}) single peak spectrum is from a Schottky contact device and independent of illumination position, thus originating from deeper, unstrained bulk phosphorus donors.}  \label{fig:curves}
\end{figure}

Additional tensor information such as the sign and orientation of the principal stress may be obtained from polarisation studies of the \dox{} spectrum, which is linearly polarised along the spin quantisation axis\cite{gullans15}, while the common shift of the \dox{} peaks can instead provide information on the hydrostatic strain component.

We next test this stress measurement approach through the scanning microscopy of donor bound excitons in microfabricated MOS Si devices (illustrated in Fig.~\ref{fig:curves}(a--d)), with phosphorus donors selectively implanted between two metal contacts.
Changes in the electrical impedance across the two terminals of the device, resulting from exciting the \dox\ transition, are measured using a lock-in and transimpedance amplifiers. The scanning optical set-up, shown in  Fig.~\ref{fig:rod}(e), achieves a laser spot size of \qty{4}{\micro\meter}, enabling local measurements of the \dox\ spectra.
The photocurrent measured as function of excitation laser energy displays a pair of resonant peaks characteristic of  a strained donor bound exciton transition (see Fig.~\ref{fig:curves}(e)). Only devices with MOS contacts yielded locally selective \dox\ photocurrent peaks consistent with locally implanted donors. Several attempts with Schottky contacts yielded a single resonant peak, independently of whether the illumination was on or off the implanted donors, thus presumably arising from the low density (\qtyrange{e11}{e12}{\per\cubic\centi\meter}) bulk  donors of the substrate. One possible explanation is that the implanted donors are already ionised in the depletion region of the nearby metal interface. In MOS devices, the measured peak splitting varies according to the device dimensions and area illuminated, ranging from tens of \unit{\micro\electronvolt} for devices with \qty{20}{\micro\meter} contact spacing to over \qty{200}{\micro\electronvolt} for a \qty{5}{\micro\meter} contact separation. The splitting is largely independent of bias voltage and consistent with the zero-field hole spin splitting induced by strain within the device.

\begin{figure}[t]
\includegraphics[width=0.99\linewidth]{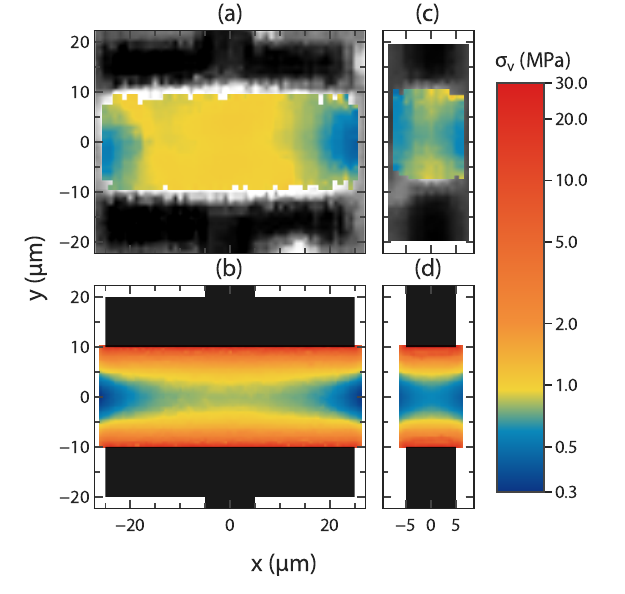}
	\caption{{\bf Equivalent stress($\sigma_v$) maps of ($20 \times 50~\upmu{\rm m}^2$) and ($20 \times 10~\upmu{\rm m}^2$) contact devices.} (a),(c) Experimental maps derived from the \dox{} peak splitting. The stress is plotted in colour where a pair of peaks are detected, while the photocurrent amplitude is plotted where they are not. (b),(d) FEM stress maps of the experimental devices.}  \label{fig:maps}
\end{figure}

Having established a technique to locally measure the \dox{} spectrum, we scan the laser spot across the device to obtain an image showing zero-field peak splitting in the region between the two metallic contacts. We scan over each device with a step-size of \qty{1}{\micro\meter}, and the spectrum obtained at each position is fit to a pair of Lorentzian functions generating a map of \dox{} peak splitting. Finally, Eq.~\ref{eq:stress_conversion} is used to convert this into an equivalent stress map, as shown in Fig.~\ref{fig:maps}(a,c).

To compare and validate the stress maps obtained, we performed a FEM simulation of thermal strain using the COMSOL solid mechanics package. The thermal expansion coefficients from silicon~\cite{watanabe04,middelmann15}, aluminium~\cite{kroeger77,wilson41} and silicon dioxide~\cite{white73,hahn72} are integrated to estimate the thermal expansion between the \qty{3}{\kelvin} experimental temperature and the fabrication temperatures (\qty{950}{\degreeCelsius} for $\mathrm{SiO_2}$ growth and \qty{300}{\degreeCelsius} for aluminium thermal evaporation).

The experimentally obtained stress images show a good quantitative agreement with the FEM maps, particularly along the $x$-axis equidistant from each contact. Along the $y$-axis, the experimental maps indicate a more uniform region of stress in the centre of the device than predicted by the FEM simulations. However, there are positions close to the contacts at which no peaks can be resolved --- these are transparent in   
Fig.~\ref{fig:maps}(a,c) --- and yet a photocurrent is observed (shown in white on the grey-scale), suggesting the presence of donors. These suggest positions where the equivalent stress is greater than  3~MPa and hence beyond scanned photon energy range(\qty{100}{\micro\electronvolt}) or in which a large stress gradient broadens the line to the point it cannot be resolved. The large peak values of stress (up to \qty{30}{\mega\pascal}, equivalent to a peak splitting of \qty{800}{\micro\electronvolt}), as well as the stress gradients (up to 15~MPa/$\upmu$m), predicted by the FEM simulations are consistent with the inability to resolve spectra at these positions adjacent to the contact edge. Piezo-scanning lasers with a wider wavelength range and decreasing the laser spot size could respectively be used to extend the applicability of this technique to larger stresses and gradients.

\section{Conclusions}
We have demonstrated the detection of donor bound excitons in microfabricated devices and characterised their local stress profiles. The strain mapping technique demonstrated appears remarkably precise at detecting local stress variations in the low strain regime, with the maps in Figure \ref{fig:maps}a,c displaying an estimated stress range as small \qtyrange{0.3}{1.3}{\mega\pascal}. The limitations in higher stress regions appear to be due to a combination of peak broadening and the limited scanning range of the fibre laser used in this work. Figure \ref{fig:curves} shows that more highly strained donors can indeed be detected where the gradient is lowered by geometric symmetry, with the greatest peak splittings corresponding to an estimated \qty{7}{\mega\pascal} and \num{e-4} strain. Our results indicate that, for devices exploiting \dox\ readout, contact separations of 10--20~$\upmu$m yield peaks which are split by strain but show minimal broadening in natural silicon, while smaller contact separations lead to large peak broadening which would interfere with spin-to-charge conversion schemes within a purely diffraction-limited geometry.
Finally, we note that the stress microscopy technique presented in this work for silicon could potentially be extended to other indirect bandgap semiconductors. For example, germanium, which is of interest for hole based quantum technologies~\cite{scappucci20,hendrickx21}, displays sharp zero-phonon \dox{} transitions~\cite{mayer79,mayer79a,davies93} and a dominant Auger recombination~\cite{klingenstein79}.  

\section{Acknowledgements}
We thank the UK National Ion Beam Centre (UKNIBC) which performed the phosphorus ion implantation and the Southampton Nanofabrication Centre  where the silicon oxide in these devices was grown.
This work has received funding from the UK Engineering and Physical Sciences Research Council (EPSRC), through the Hub in Quantum Computing and Simulation (Grant No. EP/T001062/1) and a Doctoral Training Grant (P. C.); as well as from the European Research Council (ERC) under the European Union’s Horizon 2020 research and innovation programme [Grant Agreement No. 771493 (LOQO-MOTIONS)].

\section{Methods}
\subsection{Bulk crystal measurements using ESR detection}

Stress is applied to the bulk $\mathrm{^{28}Si}$ crystal (dimensions $2\times2\times7~\mathrm{mm^3}$, doped with phosphorus at a concentration of \qty{1.5e14}{\per\cubic\centi\meter}) via a pair of machined PEEK rods which enclosing the ends of the sample. The lower rod has a circular recess to accommodate the sample, and is screwed into to an aluminium base which sits at the bottom of the cryostat. The top rod extends out of the cryostat, while an acrylic holder with a square recess is glued to the bottom of the top rod to enable rotation of the sample orientation with respect to the horizontal magnetic field, $B_0$. 
Calibrated masses are added to a platform mounted on the top rod, ouside the cryostat, to apply uniaxial compression stress to the sample, following a method used in Ref.~\cite{mansir18}

ESR detection is performed using a custom-built spectrometer~\cite{ross17} and a Bruker Flexline EN 4118X-MD4W1 resonator. The sample, sitting at \qty{4.2}{\kelvin} inside an Oxford Instruments CF935 helium flow cryostat, is illuminated through the cryostat window using a \qty{200}{\milli\second} pulse from a \qty{1078}{\nano\meter} Koheras Adjustik tunable laser, followed by a Hahn echo sequence for polarisation detection and a \qty{20}{\milli\second} above-bandgap (\qty{1047}{\nano\meter}) laser pulse to reset the spin polarisation.)

\subsection{Scanning microscopy measurements on microfabricated Si devices}

The Fz-Si devices studied consist of lithographically defined, thermally evaporated \qty{30}{\nano\meter}  thick aluminium contacts on top of \qty{11}{\nano\meter} of silicon oxide thermally grown at $950^\circ$ in an $\mathrm{O_2}$ atmosphere. Phosphorus donors are selectively implanted between the metal contacts by ion implantation with a \qty{7e10}{\per\square\centi\meter} dose at a depth of \qty{150}{\nano\meter}, resulting in a peak concentration of \qty{5e15}{\per\cubic\centi\meter}. 
The sample is fixed on a PCB using silver paint, and electrically connected by aluminium wedge wire bonding. 
A SR830 lock-in amplifier (frequency \qty{99.77}{\kilo\hertz }) and femto DLPCA-200 transimpedance amplifier (gain of \qty{e6}{\volt\per\ampere}) are used to measure changes in the electrical impedance of the two-terminal device resulting from exciting the \dox\ transition. 

The device is illuminated using a Toptica DL-Pro piezo tunable laser focused on the surface by a $\mathrm{0.35NA}$ objective, with an estimated laser spot size of \qty{4}{\micro\meter}. The device is moved under the laser spot by a 3-axis ANPx101-ANPz101 Attocube piezo-actuator. The relative position of the laser spot is determined by convolving the device image at each point with a reference image, such as the one in Fig.~\ref{fig:curves}(c), using the OpenCV function  \emph{matchTemplate}()~\cite{bradski00}. The scanning microscopy can thus be performed accurately and automatically using a custom software feedback loop between the piezo-actuator and the CMOS camera.
The laser shutter is pulse-modulated at \qty{\sim 50}{\hertz} with \qty{10}{\milli\second} pulses using a SR475 mechanical shutter 
to better isolate the photo-induced current from the background.
A \qty{1050}{\nano\meter} LED is employed outside the cryostat window to slightly increase the carrier density of the silicon substrate, providing a competing source of donor neutralisation and improving the resonant photocurrent signal. The LED is also used to provide illumination for the wide-field images shown in Fig.~\ref{fig:curves} and employed for automated device positioning. 

\subsection{Equivalent stress from spectral peak splitting}

The zero-field \dox{} peak splitting from Equation \ref{eq:splitting} closely resembles Von Mises equivalent stress $\sigma_v$.
\begin{align} \label{eq:von_mises}
\begin{split}
 \sigma_v = & \Big\{ \frac{1}{2} \big[  (\sigma_{xx}-\sigma_{yy})^2+(\sigma_{yy}-\sigma_{zz})^2 + (\sigma_{zz}-\sigma_{xx})^2  \\ & + 6(\sigma_{xy}^2+\sigma_{yz}^2+\sigma_{zx}^2 ) \big]\Big\}^\frac{1}{2} 
\end{split}
\end{align}
The experimental Pikus-Bir parameters however lead to a small difference between the coefficient for principal and shear stress. We average the two to derive a direct relationship between peak splitting and equivalent stress $\sigma_v \approx C \Delta E$:
\begin{equation} \label{eq:stress_conversion}
    \sigma_v \approx \frac{3^\frac{1}{4}}{\sqrt{2b'd'(s_{11}-s_{12})s_{44}}} \Delta E
\end{equation}
The error from this approximation is as small as 3\%, with the total coefficient error(9\%) still dominated by the  errors in the extraction of $b^\prime$ and $d^\prime$ parameters from experimental measurements.

\bibliographystyle{ieeetr}
\bibliography{strain_microscopy}

\end{document}